%% file: main.tex
\documentclass{acmart}
\usepackage{hyperref}
\usepackage{amsmath,amsfonts}
\usepackage{graphicx,color} 
\usepackage[inline]{enumitem} 
\PassOptionsToPackage{hyphens}{url} 
\usepackage{url}
\usepackage{xcolor}
\usepackage{color,colortbl}
\usepackage{tcolorbox}
\input{custom_commands}

\begin{document}

\title{Two Types of Data Privacy Controls}
\author{Eman Alashwali}
\affiliation{%
  \institution{King Abdulaziz University (KAU)}
  \city{Jeddah}
  \country{Saudi Arabia}
  }
\email{ealashwali@kau.edu.sa}
\begin{tcolorbox}
This document is the author's manuscript for an article accepted for publication at the Communication of the ACM, 2025.
\end{tcolorbox}

\maketitle

Users share a vast amount of data while using web and mobile applications. Most service providers such as email and social media providers provide users with privacy controls, which aim to give users the means to control what, how, when, and with whom, users share data. Nevertheless, it is not uncommon to hear users say that they feel they have lost control over their data on the web. 

This article aims to shed light on the often overlooked difference between two main types of privacy from a control perspective: privacy between a user and other users, and privacy between a user and institutions. We argue why this difference is important and what we need to do from here.

\section{Two Types of Privacy}
Raynes-Goldie coined the term \textit{social privacy} as opposed to \textit{institutional privacy}~\cite{raynes2010}. The former is about controlling access to personal information while the latter is about controlling how institutions such as Facebook and their partners might use this information. Heyman et al. defined the term \textit{privacy as subject} to refer to controlling a user’s personal information disclosure to other users, and \textit{privacy as object} to refer to controlling information disclosure to third parties, which represent the user as an object in a data mining process ~\cite{heyman14}. Brandimarte et al. classified privacy controls according to purpose, where \textit{release} controls refer to controlling information disclosure between users, while \textit{usage} controls refer to controlling the use of users’ information, e.g. by the service providers or third parties ~\cite{brandimarte13}. Bazarova and Masur introduced multiple approaches to privacy, which include the \textit{networked approach} where information flows in a horizontal direction between users, and the \textit{institutional approach} where information flows in a vertical direction between a user and institution ~\cite{bazarova20}. 

Here we will use the terms: \textit{user-to-user} privacy and \textit{user-to-institution} privacy. In \textit{user-to-user}, the other users could be family, friends, coworkers, etc. In \textit{user-to-institution}, the institution could be a service provider, organization, government, etc. 

\begin{tcolorbox}
Ignoring the difference between the two types of privacy controls may lead users to have an illusory sense of control over their privacy~\cite{brandimarte13,heyman14}.
\end{tcolorbox}
In recent years, many service providers, e.g. social media platforms, have improved the privacy controls provided to users. However, they may have improved one type of privacy controls: the \textit{user-to-user}~\cite{heyman14}. Ignoring the difference between the two types of privacy controls may lead users to have an illusory sense of control over their privacy~\cite{brandimarte13,heyman14}. For example, users’ perceived control over \textit{user-to-user} privacy may result in fewer privacy concerns as a result of an incomplete assessment of the associated risks of data sharing, ignoring what Stutzman called \quotes{silent listeners}~\cite{stutzman13}. The \textit{user-to-user} privacy is a subset of the \textit{user-to-institution} privacy. However, when service providers emphasize the \textit{user-to-user} privacy controls, or even worse, do not offer \textit{user-to-institution} controls, users can get confused about where did they fail in controlling their privacy.

The distinction between the types of privacy is also important in privacy surveys and studies. Looking at privacy constructs, such as awareness, behaviors, attitudes, and concerns, researchers need to design their surveys and studies with the distinction in mind to avoid any confusion. They should clearly communicate which type of privacy controls the study or question is about. Let’s look at privacy awareness for example: users may be aware of the more prominent \textit{user-to-user} privacy controls such as restricting Facebook’s profile information visibility from other users, but not the \textit{user-to-institution} controls such as restricting what information the provider can use in the ads shown to the user. Similarly, for privacy concerns, users may be more concerned about the \textit{user-to-user} privacy (e.g. that their manager sees a Facebook post they did not want them to see) than an advertiser uses their post information to show relevant ads, or vice versa.

\begin{tcolorbox}
The distinction between the types of privacy controls is important to account for multi-cultural differences in privacy perceptions that may arise in one type of privacy controls, but not the other.
\end{tcolorbox}

The distinction between the types of privacy controls is important to account for multi-cultural differences in privacy perceptions that may arise in one type of privacy controls, but not the other. For example, we may find different privacy awareness and behaviors for the \textit{user-to-user} privacy controls due to societal norms that drive a society to be more aware of the \textit{user-to-user} privacy. For example, in conservative societies, a considerable fraction of women do not prefer to share their photos publicly, thus, they are well-aware of how to hide their profile visibility on social media platforms. On the other hand, these cultural differences may not appear in the \textit{user-to-institution} privacy controls, as the societal norms are less relevant here.  

Drawing a line between the two types of privacy in privacy studies, surveys, and discussions is important for an accurate perception and understanding of the privacy issues societies face.

\begin{tcolorbox}
As researchers, we need to agree on accurate and sensible terms to describe the different types of data privacy from a control perspective.
\end{tcolorbox}
\section{What Do We Need to Do Next?}
From here, first, as researchers, we need to agree on accurate and sensible terms to describe the different types of data privacy from a control perspective. To this end, we first need to identify what terms are already there in the literature -- we listed some of the terms we are aware of earlier in this article, but there might be more. We do not only need a list of existing terms, but an understanding of the reasoning behind them and their definitions, if any. We then need to provide definitions for each type of privacy controls in a more systematic way, including the actors and data flows that each type involve. Before we move to suggest rigorous studies to evaluate existing terms, we may need to crowdsource more terms that can capture the definitions more precisely, both from privacy experts and non-experts. From there, we can move forward to evaluate users’ comprehension and sentiments towards the terms. Eventually, we should be able to shortlist, then identify, the most sensible terms that accurately define the two types of privacy controls. To realize this, we will likely need a combination of both qualitative and quantitative approaches. There have been studies in the literature that examined terminology issues, for example, the terms used in cookie consent interfaces~\cite{jiwani24} and privacy policies~\cite{tang21}, which we can learn from. 

After identifying the most sensible terms to describe both types of privacy controls, researchers and the industry need to adopt them, and raise awareness about the different types of privacy controls. We should adopt sensible and common terms in our product designs, research studies, and in our privacy discussions in general. Users eventually should adopt these terms too and be more precise in communicating their privacy perceptions, behaviors, and concerns.

Reaching a consensus on terms is not going to be free from limitations. For example, from my own perspective as a bilingual, I wonder if I conducted the study for choosing the most accurate and sensible terms in English as a representative language for service providers, would the results still hold if the terms were translated to another language? This may require follow-up studies.

Finally, with precise, sensible, easy to comprehend and use terms to differentiate the two intrinsic types of privacy controls, I believe this will positively impact the accuracy of privacy research and discussions, and this is a worthwhile endeavor. 


\section{About the Author}
Eman Alashwali is an Assistant Professor at King Abdulaziz University (KAU), Jeddah, Saudi Arabia. She earned her DPhil (Phd.) in Cybersecurity from the University of Oxford, UK in 2020 and was a Collaboration Visitor at the Carnegie Mellon University’s Security and Privacy Institute (CyLab) from 2022 to 2024. She also holds a MSc. in Information Security form University College London (UCL), UK and BSc. in Computer Science from KAU.

\bibliographystyle{apalike}
\bibliography{refs}
\end{document}

%% file: custom_commands.tex
\newcommand{\quotes}[1]{``#1''} 

\newcommand{\autorefappendix}[1]{\hyperref[#1]{Appendix~\ref*{#1}}}  
